\documentclass[%
aip,
amsmath,amssymb,
reprint,%
]{revtex4-1}

\usepackage{float}
 \usepackage{chemformula}
\usepackage{graphicx}
\usepackage{xcolor}
\usepackage{bm}
\usepackage{amsmath,amssymb}
\usepackage[utf8]{inputenc}
\usepackage[version=3]{mhchem} 

\begin{document}
		\title{Theory of Charge Regulation of Colloidal Particles in Electrolyte Solutions}
	\author{Amin Bakhshandeh}
		\email{bakhshandeh.amin@gmail.com}
\affiliation
	{Instituto de F\'isica, Universidade Federal do Rio Grande do Sul, 91501-970, Porto Alegre, RS, Brazil}

	\author{Derek Frydel}
		\email{derek.frydel@usm.cl}
\affiliation
	{Department of Chemistry, Federico Santa Maria Technical University, Campus San Joaquin,7820275,  Santiago,Chile}

	\author{Yan Levin}
	\email{levin@if.ufrgs.br}
\affiliation
	{Instituto de F\'isica, Universidade Federal do Rio Grande do Sul, 91501-970, Porto Alegre, RS, Brazil}


	\begin{abstract}
		We present a theory that enables us to calculate the effective surface charge of colloidal particles and to efficiently obtain titration curves for different salt concentrations.   The theory accounts for the shift of pH of solution due to the  presence of  1:1 electrolyte.  It also accounts self-consistently for the electrostatic potential produced by the deprotonated surface groups.  To examine the accuracy of the  theory we  have performed extensive reactive Monte Carlo simulations, which show excellent agreement between theory and simulations {\color{black}without any adjustable parameters}.
	\end{abstract}
	\maketitle
	\section{Introduction}\label{introduction}
	Colloidal particles are important for various applications in chemistry, biology, and physics~\cite{butt1994technique,levin2002electrostatic,andelman2006,Borkovec2001,israelachvili2011}. The vast  variety of applications of colloidal systems has made the modern society very much dependent on these complex systems~\cite{vonarbourg,everett2007basic,coldis,hama,dickinson2007food}. 
	To stabilize colloidal suspensions, the particles are often synthesized with acidic or basic surface groups.   In aqueous suspensions these groups become ionized leading to formation of electrical double layer (EDL)~\cite{grahame1947electrical,guldbrand1984electrical,lopez2007electrical,james1982characterization,attard2001recent,carnie1994electrical,Krishnan,hiemstra1996surface,hiemstra1996intrinsic}. The complicated physics of EDL is responsible for  stability of colloidal suspensions and can lead to some very counterintuitive effects, such as reversal of electroscopic mobility in suspensions with multivalent counterions or like-charge attraction between colloidal particles with the same sign of  charge~\cite{quesada2002interaction,fernandez2005ionic,pianegonda2005charge,guerrero2010effects}. The stability of colloidal systems is usually explored using ideas first introduced by   Derjaguin, Landau, Verwey and Overbeek (DLVO)~\cite{hermansson1999dlvo,ninham1999progress,Verwey,boon2015}.   However, DLVO theory does not take into account electrostatic correlations~\cite{levin2002electrostatic} between ions and between ions and sites.  {\color{black} Such effects can become very important for suspensions containing multivalent counterions and also in suspension containing large concentrations of 1:1 electrolyte.  }
	Furthermore, the charge of colloidal particles is not constant, but
	depends on the pH and electrolyte concentration inside the suspension.  The fluctuation of colloidal charge can lead to some very non-trivial effects, in particular close to the isoelectric point~\cite{avni2018charge}.
	The process of charging of colloidal particles is denoted as charge regulation (CR) and was first elucidated by  Linderstr{\o}m-Lang~\cite{linderstrom}.  The first quantitative model of CR was proposed by Ninham and Parsegian~\cite{ninham1971}.  {\color{black} The NP approach is based on Poisson-Boltzmann (PB) theory, which neglects the discrete nature of colloidal surface groups and electrostatic correlation effects~\cite{frydel2019,pod2019,AVNI201970,Markovich2016,Gary,Curk,Madhavi1}.  Indeed PB theory is known to be very accurate for suspensions with monovalent 1:1 electrolytes.}  However recent works, based on Baxter's model of sticky spheres to describe protonation/deprotonation equilibrium, showed~\cite{bkh2021,bk2019,Bakhjpa2020,bkh2022ma} that discreteness of surface groups affects significantly the effective charge predicted by the NP theory.  {\color{black} The importance of ion  polarizability and of finite ion size was also explored in reference \cite{parsons2019thermodynamic}}.   
	
	The Baxter model of protonation also  showed that  there is a change in equilibrium constant when an acidic group is moved from bulk to the surface~\cite{bkh2021,bk2019,Bakhjpa2020,bkh2022ma}.
	{\color{black} In general, the renormalization of the equilibrium constant is  due to the
		broken rotational and translational symmetry at the surface compared to the bulk.}  The theory developed, however, showed that both the discrete charge effects and the renormalization of the bulk equilibrium constant by the surface can be included within the NP framework.  The calculations, however, were not entirely self-consistent, since the discreteness effects were not taken into account at the same level of mean-field approximation as are implicit in the PB equation.  In the present work we correct this omission and also account for the shift in pH produced by  the dissolved electrolyte.  Both of these effects become important for suspensions with weak acidic surface  groups in suspensions with salt.
	
	The rest of the paper is organized as follows: in the section ``Theory", we present the fully self-consistent theory of charge regulation.   In the section ``Reactive Monte Carlo Simulations", we briefly describe the simulation method used to compare with the predictions of the theory.   In section ``Transcendental Approximation" we present a simple transcendental equation that provides us with an easy way of calculating the effective colloidal charge and titration curves in suspensions at large dilution. Finally, in the last section we  present the conclusions of the present work.

	\section{Theory}
	
	\begin{table}[H]

	\begin{tabular}{cc}
		\hline
	Symbol & Significance    \\
	\hline
$\lambda_B$= 7.2 \AA 	& Bjerrum length \\
	\hline

	$\beta$   	&   $1/k_BT$  \\
			\hline
$f_0$   	&   free energy of a deprotonated state of a site \\
	\hline
		$f_1$   	&   free energy of a protonated state of a site  \\
		\hline	 
	$\mu$   	&  chemical potential  \\
	\hline
		$K$   	&  equilibrium constant  \\
	\hline
		$\Lambda $   	&   de Broglie thermal wavelength  \\
	\hline
			$c_i$   	&  concentration of electrolyte   \\
	\hline
				$a_i$   	&  activity of electrolyte   \\
	\hline  $\phi_s$  & isolated site electrostatic potential\\
	\hline
 			$\phi$   	&  mean-field electrostatic potential   \\
 \hline
  			$\kappa$   	&  Deby length  \\
 \hline
   		$M = 1.106$  	&  Madelung constant   \\
 \hline
    		$ \epsilon_w$  	& dielectric constant of water   \\
 \hline
     		$ a$  	& colloid radius   \\
 \hline   		
     		$\gamma$ & colloidal volume fraction\\
 \hline
     		$ R$  	& cell size  \\
 \hline
      		$ Z_{eff} $  	& number of deprotonated groups  \\
 \hline
       		$ K_a=1/K  $  	&  acid dissociation  constant  \\
 \hline
        		$ q  $  	&   proton charge  \\
 \hline
         		$\varphi({\bf r})$   	&   electrostatic potential at position ${\bf r}$  \\
 \hline                 
 			$Q$   & colloidal charge \\
      		
 \hline
          		$\sigma$   	&   surface charge density \\
 \hline
	\end{tabular}
\caption{\label{symbol}Symbols used in the rest of the paper.}
	\end{table}
	We study a colloidal particle of radius $a$, containing $Z$ acidic surface groups.  The particle is confined inside a spherical Wigner-Seitz (WS) cell of radius $R$, which is determined by the concentration $\gamma$ of colloidal suspension, $\gamma=3/4 \pi R^3$.  The suspension is in contact with a reservoir of salt and acid at concentrations $c_s$ and $c_a$, respectively. {\color{black}For reader's convenience, in Table~\ref{symbol} we present a list of symbols that appear in 
	the rest of the paper.} 
	
	\textcolor{black}{The number of deprotonated surface groups  $Z_{eff}$ is determined by the chemo-thermodynamic equilibrium.}
	The charge of the colloidal particle can then be written as  $Q=-Z_{eff}q=-Z q \xi$, where $q$ is the proton charge and $\xi$   is the probability that a surface group is deprotonated: 
	\begin{eqnarray}\label{a1}
	\xi =\frac{\exp(-\beta f_0)}{\exp(-\beta f_0) + \exp(-\beta f_1)} .
	\end{eqnarray}
	
	In this expression $f_0$ is the free energy of a deprotonated state and  $f_1$ is the free energy of a protonated state of a site. \textcolor{black}{  $f_0$ is due to electrostatic free energy of solvation of a surface group inside an electrolyte solution,  $f_0=\mu_{solv}$.   The ions of electrolyte partially screen the electric field produced by a surface group lowering its overall electrostatic self energy}.   Dividing the numerator and denominator of Eq. (\ref{a1}) by $\exp(-\beta f_0)$, we can write
	\begin{eqnarray}\label{a11}
	Z_{eff} =\frac{Z }{1 + \exp(-\beta \Delta \mu)}  
	\end{eqnarray}
	where $ \Delta \mu=f_1-f_0$  is the difference in free energy between protonated and deprotonated states of surface active groups.    \textcolor{black}{We should note that $Z_{eff}$ should not be confused with the far field effective charge often defined in studies of colloidal systems~\cite{Pincus84,aubouy2003effective,levin2002electrostatic}}
	
	Consider a reaction occurring on the colloidal surface  
	\begin{eqnarray}\label{K}
	\ce{H+}+\ce{A-} \rightleftharpoons \ce{HA},\\ \nonumber
	\end{eqnarray}
	where   \ce{A-} is a surface deprotonated group and $K$ is the equilibrium constant of the reaction.  The equilibrium constant $K$ is related to the weak acid dissociation constant by $K=1/K_a$.
	As discussed in the introduction, the value of  surface $K$ is in general different from the same reaction occurring in the bulk.  $K$ accounts for the direct electrostatic interaction of proton with an isolated surface group bound to the surface.  This can only be calculated using quantum density functional theory.  At the semiclassical level, we can denote  $\zeta=K/\Lambda_{ \ch{H+}}^3$ --- where $\Lambda_{ \ch{H+}}$ is the proton de Broglie thermal wavelength ---  as the internal partition function for a $\ch{HA}$ ``surface molecule".  In addition to this interaction, when proton is moved from the reservoir to colloidal surface it will also interact with the other surface groups.  Since, these groups are reasonably far away, the quantum effects can be neglected and the long range interaction can be modeled using classical electrostatics.   The change in free energy  due to removal of a hydronium ion from the reservoir and transferring it to colloidal surface, where the reaction takes place is then:
	\begin{equation}\label{k3}
	\beta \Delta \mu =-\ln( K /\Lambda_{ \ch{H+}}^3)+\beta q \varphi- \beta \mu_{solv}-\ln{(c_a \Lambda_{ \ch{H+}}^3)}-\beta \mu_{ex}\,.
	\end{equation} 
	The first term on the right hand side of this expression is the free energy of direct interaction of proton with the adsorption site, the second term is the mean electrostatic energy of proton interacting with all the other deprotonated acid groups and with the ions of solution.  The third term is the loss of electrostatic solvation free energy when the site becomes protonated (neutral).  Finally, the last two terms are the free energy change of the reservoir when one hydronium ion is moved to the colloidal surface. {\color{black} The excess chemical potential $\mu_{ex}$ is an important part of the proton activity, $a_{\ch{H+}}=c_a e^{\beta \mu^{ex}}/c_a^\circ$, where $c_a^\circ=1$M is the standard reference concentration.     
		The pH of a suspension containing electrolyte is defined as pH$=-\log_{10}[a_{\ch{H+}}]$}.  For 1:1 electrolyte  $\mu_{ex}$  is very accurately accounted for using the mean spherical (MSA)~\cite{ho1988mean} and Carnahan-Starling (CS) approximations, $\mu^{ex} = \mu_{CS} +\mu_{MSA}$,  where
	\begin{equation}\label{msa}
	\beta \mu_{MSA}= \frac{\kappa d \sqrt{1 + 2 \kappa d}- \kappa d -(\kappa d)^2 }{8
		\pi c_t d^3}
	\end{equation}
	and 
	\begin{equation}\label{cs}
	\beta \mu_{CS}=\frac{8\eta-9 \eta^2+3\eta^3}{\left(1-\eta\right)^3} 
	\end{equation} 
	where $d=4$~\AA\ is the ionic  diameter, which for simplicity we take to be the same for all ions, $\kappa=\sqrt{8 \pi \lambda_B c_t}$ is the inverse Debye length,  $c_t=c_a+c_s$, 
	$\lambda_B=q^2/k_B T\epsilon_w$ is the Bjerrum length which is $7.2$~\AA\ in water at room temperature, and  $\eta=\frac{\pi d^3}{3} c_t$.  Substituting this into Eq. (\ref{a11}), we obtain
	\begin{equation} \label{zeff}
	Z_{eff}=\frac{Z}{1+ K c_a e^{-\beta(\varphi -\mu^{ex}-\mu_{sol})}}
	\end{equation} 
	
	\subsection{Electrostatic free energy of solvation of surface site}
	To calculate the  electrostatic solvation free energy of an isolated group of charge $q_s$ located on the surface of a colloidal particle, we neglect the curvature effects and threat the surface as an infinite plane.  This is very reasonable for suspensions containing a lot of salt  --  for which electrostatic solvation energy is significant  --  since the electrostatic curvature effects will be screened on the scale larger than the Debye length.  
	
\begin{figure}[H]
		\centering
		\includegraphics[width=0.4\linewidth]{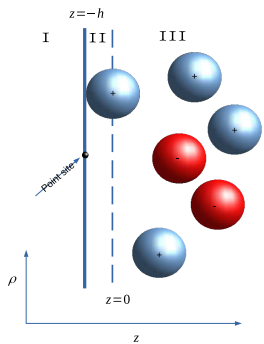}
		\caption{Schematic representation of an isolated adsorption site on colloidal surface, with exclusion zone due to hard core repulsion of ions from the surface.}
		\label{figure}
	\end{figure}

	We will work in cylindrical coordinate system, with the colloidal surface located at $z=-h$,  where $h=d/2=r_{ion}$, see Fig.~\ref{figure}.  Because of hardcore repulsion between ions and colloidal surface, we see that there is an exclusion layer where no ions are present.  Within the Debye-H\"uckel approximation, the electrostatic potential then satisfies
	\begin{eqnarray}\label{g2}
	\left\{
	\begin{array}{ll}
	\nabla^2  {\phi_s}(\rho,z) =- 2 q_s \frac{\delta(\rho) \delta(z+h)}{\rho}, &z<0 \\
	\nabla^2  {\phi_s}(\rho,z) = \kappa^2   {\phi_s}(\rho,z), & 
	z>0
	\end{array}
	\right.
	\end{eqnarray}
	Using the azimuthal symmetry of the problem, the solution can be written as~\cite{Levin_2001} :
	\begin{eqnarray}\label{g3}
	\phi_s(\rho,z) = \frac{q_s}{2\pi}\int_0^\infty dkkJ_0(k\rho)\tilde{\phi_s}(k,z)
	\end{eqnarray}  
	Substituting this into Eqs. (\ref{g2}) we obtain
	
	
	
	
	\begin{eqnarray}\label{g15}
	\tilde{\phi_s}(k,z) = \frac{2 \pi }{k}\left[e^{-k|z+h|}+\frac{k-p}{k+p}e^{k(z-h)}\right] \text{for}\,\, z<0,
	\end{eqnarray}
	where $p=\sqrt{k^2 + \kappa^2}$.
	Substituting Eq. (\ref{g15}) back into Eq. (\ref{g3}), the integral over the first term can be performed analytically, resulting in the usual Coulomb potential produced by the charge $q_s$ located at $(\rho=0, z=-h)$.  The second term, therefore, gives us the induced potential produced by the polarization of the ionic atmosphere by the surface charge group.  The electrolyte partially screens the electric field of the charged site, resulting in negative electrostatic solvation free energy, which can be calculated using the G\"untelberg  charging process~\cite{guntelberg1926, Levin_2001}.  We find~\cite{gomez2021lattice}
	\begin{equation}\label{sol}
	\beta \mu_{sol}=\frac{\lambda_B}{2}\int_0^\infty\frac{k-\sqrt{\kappa^2 +k^2}}{k+\sqrt{\kappa^2 +k^2}} e^{-2 k r_{ion}}dk\,.
	\end{equation}
	
	\section{Electrostatic potential} 
	
	The electrostatic potential that an adsorbed  proton feels can be separated into two contributions: the direct electrostatic interaction with the adsorption site and the interaction with the other sites and with the ions of electrolyte.  The direct interaction with the adsorption site is already included inside the $\Delta \mu$ through the equilibrium association constant $K$.  The interaction with the other sites and with the ions of electrolyte, $\varphi$, can be separated into two contributions by adding and subtracting a uniform neutralizing background, $\varphi=\phi_0+\phi_{disc}$, where $\phi_{0}$ is the potential produced by the subtracted neutralizing background together with the ions of electrolyte.  
	This potential is very close to the mean surface potential produced by a {\it uniformly} charged sphere, of charge density $\sigma=-Z_{eff}q/4\pi a^2$, inside an electrolyte solution.   The potential $\phi_{disc}$ is then, the electrostatic potential produced at the position of adsorption site $i$ by the other deprotonated sites and by their neutralizing background.  
	
	We will assume that the sites on the surface of colloidal particle have hexagonal order.  Strictly speaking one can not tile a spherical surface with hexagons,  so defects must be present.  The defects, however, modify only slightly the electrostatic energy~\cite{levin2003charges} .  The optimum distribution of charges on a spherical surface, such that it minimizes the electrostatic Coulomb energy, is a well studied Thomson ordering problem~\cite{Thomson,perez1997influence,bowick2002,Bausch}.  The electrostatic energy of $Z$ point sites of charge $-q$,  arranged with pseudo-hexagonal Thomson order on a surface of a sphere with a uniform neutralizing background is very well approximated by~\cite{levin2003charges} 
	\begin{equation}\label{EOCP}
	E_{dis}  \approx  \frac{-M q^2 Z^{3/2}}{2 a \epsilon_w}  
	\end{equation}
	where $M = 1.106$ is the Madelung constant of a planar hexagonal lattice of charges on a neutralizing background.  
	If $n$ of the surface sites are protonated,  the mean charge of each surface site is $q(1-\frac{n}{Z})$.  In equilibrium protons can hop between the sites, so that at the mean-field level of approximation the energy of a Thomson sphere with $n$ neutralized sites is:	
	\begin{equation}\label{EOCP1}
	E_{disc}(n) = \frac{-M  Z^{3/2}q^2\left(1-\frac{n}{Z}\right)^2}{2 a \epsilon_w} .
	\end{equation}
	Note that at the mean-field level of approximation, we neglect the correlations between the condensed protons so that their charge is effectively smeared uniformly between the adsorption sites.  This is precisely the same level of approximation that is implicit in the Poisson-Boltzmann equation, which also neglects the electrostatic correlations~\cite{levin2002electrostatic}.  
	
	The $\phi_{disc}$ is the change in energy when and additional site becomes protonated $\Delta n=1$,
	\begin{equation}\label{EOCP2}
	\beta \phi_{disc}   \approx \frac{\partial E_{disc}}{\partial n}=\frac{\lambda_b M \left(1 -\frac{ n}{Z}\right)\sqrt{Z}}{a \epsilon_w} =\frac{\lambda_b M Z_{eff}}{a \epsilon_w \sqrt{Z}}
	\end{equation}

\begin{figure}[H]
		\centering
		\includegraphics[width=0.7\linewidth]{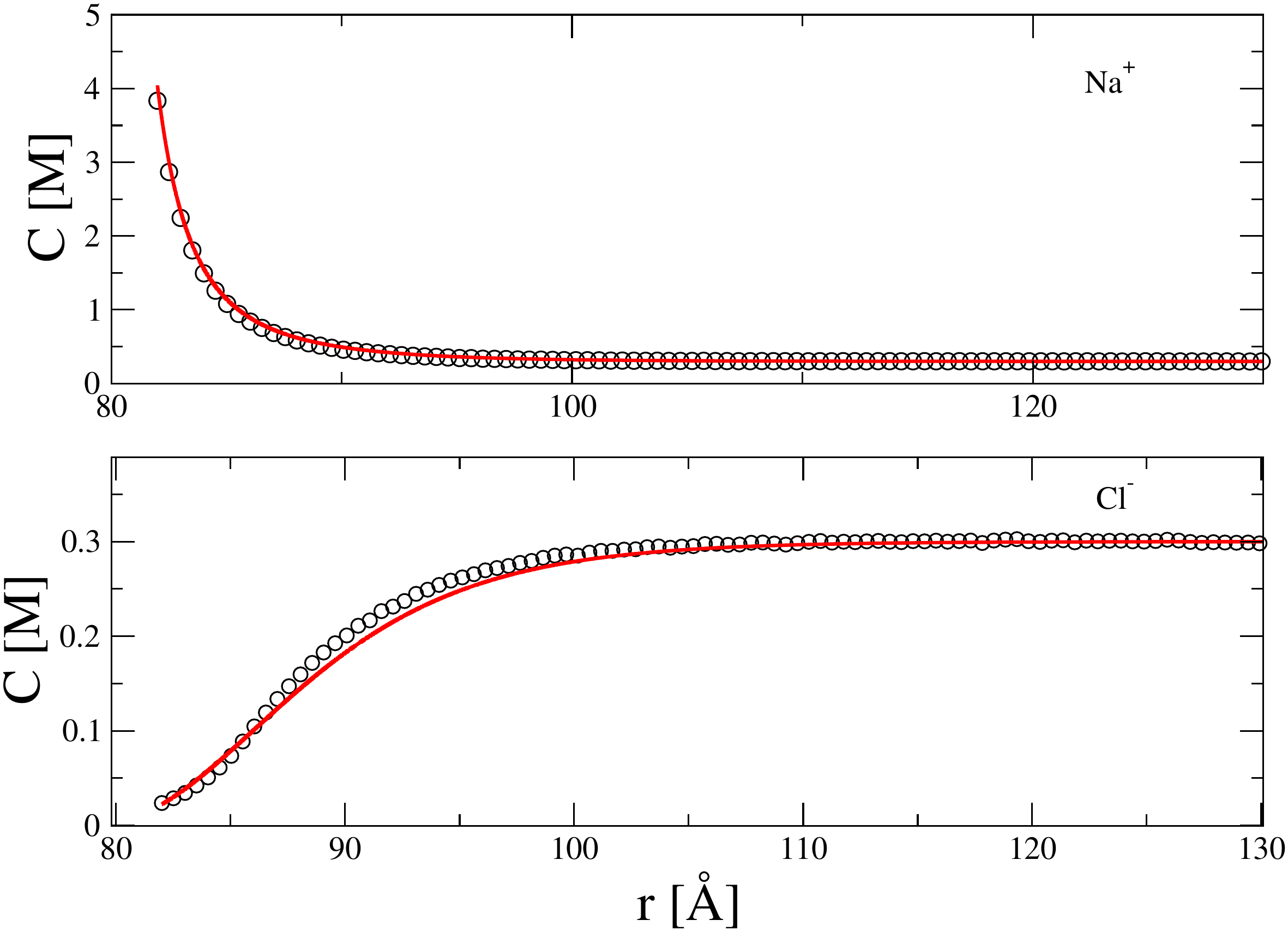}
		\caption{Comparison of  ionic density profiles obtained using simulation (symbols) and PB equation for a colloidal particle of  radius 80~\AA ~and 600 point charged site randomly distributed on its surface, inside an electrolyte  solution of 300 mM.}
		\label{fix}
	\end{figure}

	The concentration of ions inside a WS cell is determined by the equivalence of electrochemical potentials in the reservoir and in the WS cell
	\begin{equation}\label{chem}
	q_i \varphi({\bf r}) + \ln[c_i({\bf r})] + \mu^{ex}({\bf r}) = \ln(c_i^{res} ) + \mu^{ex}\,
	\end{equation}
	where $\varphi({\bf r})$ is the electrostatic potential at position ${\bf r}$ and $c_i$ is  concentration of ion of type $i$.  
	We will suppose that the excess contribution to the chemical potential in the reservoir and inside the WS are approximately the same and will cancel out.  Furthermore, the discreteness effects of the surface groups decay very rapidly away  from colloidal surface, so that  they  can be replaced by a uniform surface charge density, so that $\varphi({\bf r})=\phi(r)$ for $r>a+r_{ion}$.  Combining this with Eq. (\ref{chem}), we arrive at the usual Poisson-Boltzmann (PB) equation.
	\begin{equation}\label{mpb}
	\nabla^2 \phi(r) =  \frac{8 \pi q }{\epsilon_w} \left(c_a+c_s\right) \sinh[\beta \phi(r)] ,
	\end{equation}   
	Since the hardcore repulsion will prevent presence of ions within $a<r<a+r_{ion}$, in this region the electrostatic potential will satisfy the Laplace equation.  To check the validity of our ``smearing'' approximation in which the discrete surface charge is replaced by a uniform surface charge density $\sigma=-Z_{eff}q/4\pi a^2$ in order to calculate the ionic distribution, we perform a grand canonical Monte Carlo (GCMC) to obtain the ionic density profiles inside the WS cell for a colloidal particle with  $Z_{eff}=600$ charged sites.  We then compare this profiles to the ones calculated using Eq. (\ref{mpb}), see
	Fig.~\ref{fix}.
	\vspace{1cm}
	We see a very good agreement between ionic density profiles calculated using GCMC simulations and PB equation.  As can be observed, the discreteness effects are not important away from colloidal surface. The lack of correlational effects on ionic distribution can be partially attributed to the cancellation of the correlational contribution to the electrochemical potential between the reservoir and the system, as discussed following Eq. (\ref{chem}).  We can now identify the value of $\phi(a)$ with  $\phi_0$.   
	
	In the calculation above we have arbitrary fixed the effective charge, in reality it must be determined self-consistently using Eqs. (\ref{zeff})  and (\ref{mpb}).  To solve these equations we proceed iteratively.  We first guess the mean-field potential $\phi(a)=\phi_0$.   For this guess we solve Eq. (\ref{zeff}) to numerically determine the colloidal charge.   The Gauss law then provides us with the electric field, or equivalently $\phi'(a)=Z_{eff}q/\epsilon_w a^2$.  Using $\phi(a)$ and $\phi'(a)$ as initial conditions, we then integrate the PB equation (\ref{mpb}) using the Runge-Kutta 4th order algorithm.  If the electric field at the cell boundary $r=R$, or equivalently $\phi'(R)$,  is not zero, as is required by the overall charge neutrality of the system, we adjust our initial guess for $\phi_0$.  In practice, finding the correct surface potential $\phi_0$ is facilitated by combining the algorithm described above with the Newton-Raphson root finding subroutine.  
	
	\section{Reactive Monte Carlo Simulations} 
	
	There are different simulation methods available to calculate the effective charge in CR systems. {\color{black} To the best of our knowledge, the first Monte Carlo simulation method for titration was introduced by Nishio~\cite{nishio1994monte} in 1994.   The subsequent research extended this early work to account for the over all charge neutrality inside the simulation cell and for the presence of explicit hydronium ions~\cite{lund11,Frezza,Teixeira,lund2005enhanced,landsgesell2019simulations, landsgesell2020grand,stornes2021polyelectrolyte}.   Here we will use a reactive Monte Carlo (rMC) simulation method~\cite{bakhshandeh2022reactive}, which is particularly easy to implement for the present colloidal system. Just like in the theory described above, the colloidal particle is located at the center of a Wigner-Seitz (WS) cell, which is in  contact with an infinite reservoir of salt and strong acid. 
		The WS cell radius is $R=120$~\AA\ (unless specified differently) and colloidal radius is 80~\AA.   The intrinsic pK$_a=-\log_{10} K_a=\log_{10} K$ of a functional group on the colloidal surface is taken to be $5.4$, similar to that of carboxylic acids.}
	The functional groups are treated as point sites  located on the colloidal surface. The reservoir contains strong acid, \ch{HCl}, and strong electrolyte \ch{NaCl},  which are assumed to be fully dissociated.  A proton associates with water molecule forming a hydronium ion.  {\color{black}  Again to be consistent with the theory above, we will treat all ions as having the same radius $r_{ion}=2$~\AA.  Water is treated implicitly, with Bjerrum length set to $\lambda_B=7.2$~\AA.  The interaction energy between all particles includes the normal Coulomb potential and a hardcore repulsion between ions, colloidal particle, and WS cell boundary.}
	
	{\color{black} The simulation consists of standard grand canonical Monte Carlo (GCMC) insertion/deletion moves, as well as protonation and deprotonation moves. The insertion/deletions and protonation/deprotonation moves must always involve cation-anion pair to preserve the charge neutrality inside the simulation cell. 
		The excess chemical potential $\mu_{ex}$ inside the reservoir can be calculated using a separate simulation.  This can be done using Widom insertion method in a canonical MC with fixed concentration of acid and salt in a cubic simulation cell with periodic boundary conditions, or using a reverse GCMC strategy in which the value of $\mu_{ex}$ in GCMC is adjusted until the target concentration inside the simulation cell is reached~\cite{Widom,bakhshandeh2022reactive}.  In order to accurately calculate the excess chemical potential for a specific concentration of acid and salt inside an {\it infinite reservoir} it is important to use Ewald summation to treat all the electrostatic interactions between ions.}
	
	\begin{figure}[H]
		\centering
		\includegraphics[width=0.4\linewidth]{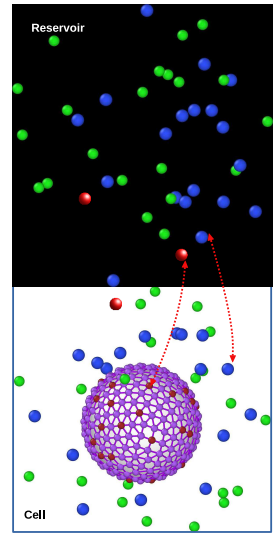}
		\caption{{\color{black}Schematic representation of reactive MC moves. Red, blue, and green spheres are \ch{H3O+}, Cl$^-$ and Na$^+$, respectively. The pair \ch{H3O+} and \ch{Cl-} enters the cell from reservoir, H$^+$ can go to the colloidal surface and react with a site, while Cl$^-$ moves into the bulk with rMC probabilities given by Eqns.(\ref{eq7}).   Alternatively \ch{H3O+} can also move to the bulk.  Similarly a pair  \ch{Na+} and \ch{Cl-} can move between the cell and the reservoir.  All the bulk moves are performed with standard GCMC probabilities. All moves are done in pairs to preserve the charge neutrality inside the simulation cell. }}
		\label{de}
	\end{figure}

	The acceptance of titration move is given by 
	$acc \rightarrow \min\left(1,\phi_{p/d}\right)$, where $p$ refers to protonation and $d$ to deprotonation~\cite{bakhshandeh2022reactive}:
	\begin{equation}\label{eq7}
	\begin{split}
	\phi_p = \frac{ c_{\ch{H+}}K  V c_{\ch{Cl-}}}{(N_{\ch{Cl-} } +1)}\exp\left[-\beta \left(\Delta E_{ele} -\mu^{ex}_{\ch{H+}} - \mu^{ex}_{\ch{Cl-} }\right)\right] ,\\
	\phi_d = \frac{ N_{\ch{Cl-}}}{c_{\ch{H+}}K V c_{\ch{Cl-}}}\exp\left[-\beta \left(\Delta E_{ele} +\mu^{ex} _{\ch{H+} } + \mu^{ex} _{\ch{Cl-} }      \right)\right]
	\end{split}
	\end{equation}
	where $N_{\ch{Cl-}}$, $V$, $\Delta E_{ele}$ are:  the number chloride ions, accessible volume of the WS cell, and change in electrostatic energy inside the cell upon a trial move, respectively.  It is important to note that titration move is always combined with insertion or deletion of \ch{Cl-} to preserve the charge neutrality of the system. See the schematic in Fig. \ref{de}. The $\mu^{ex} _{\ch{H+}}$ and $ \mu^{ex} _{\ch{Cl-}}$ are the excess chemical potentials of hydronium and of \ch{Cl-} in the reservoir.  In the present model, with all ions of the same size, $ \mu^{ex} _{\ch{H+}}= \mu^{ex} _{\ch{Cl-}}=\mu^{ex} _{\ch{Na+}}$.  The same value of $K_a=1/K=3.95 \times 10^{-6}$M, corresponding to carboxylic acid,  is used in simulations and in the theory, so that there are no adjustable parameters.
	
\begin{figure}[H]
		\centering
		\includegraphics[width=0.7\linewidth]{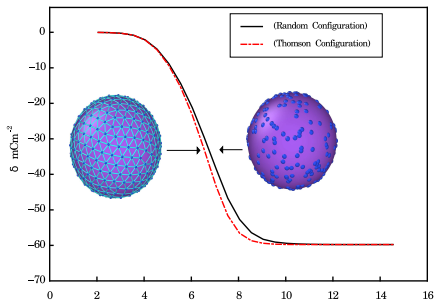}
		\caption{Comparison of titration curves for Thomson and random distributions of surface charge groups. }
		\label{fig1}
	\end{figure}

	We start by studying the dependence of titration curves on the distribution of surface charge groups.  Two possibilities are explored -- (1) a random distribution of sites and (2) annealed  distribution in which sites are first allowed to arrange on the surface of a sphere, so as to minimize their repulsive Coulomb energy.  This results in Thomson, pseudo-hexagonal, ordering of sites on colloidal surface.  
	The Thomson configuration of reactive sites is then frozen and rMC is performed.  We recall that pH$=-\log_{10}  a_{\ch{H+}}$, where $a_{\ch{H+}}=e^{\beta \mu_{\ch{H+}}}/\Lambda^3_{\ch{H+}} c^\circ$, so that  pH can be written as pH$=-\log_{10} [\ch{H+}/c^\circ] -0.434294 ~\beta \mu^{ex}_{\ch{H+}}$.
	In Fig. \ref{fig1} we see that there is some dependence of titration curves on the distribution of adsorption sites.  To be consistent with the theoretical model, and to avoid calculating averages over disorder,  in the rest of this paper we will use Thomson site distribution.

\begin{figure}[H]
		\centering
		\includegraphics[width=0.7\linewidth]{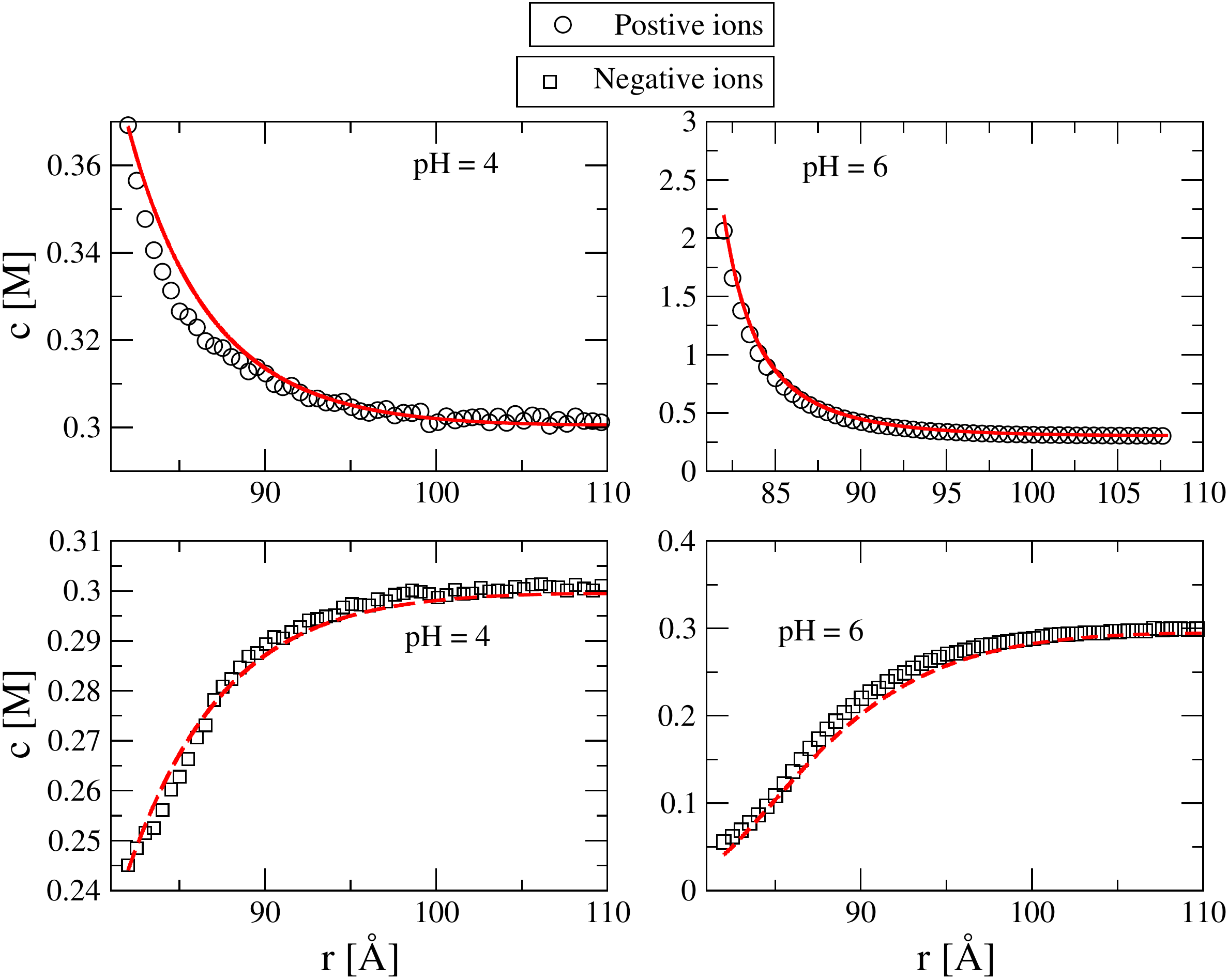}
		\caption{Comparison of ionic density profile for  pHs 4 and 6 of a colloidal particle with  600 active sites with  $K_a=1/K=3.95 \times 10^{-6}$ M and radius 80~\AA\, the concentration of 1:1 salt is 300 mM.  Symbols are simulation results and solid curve is the theory.}
		\label{fgd}
	\end{figure}

	We now apply the theory developed above to calculate  the ionic density profiles around 
	colloidal particle of radius $80$~\AA\ with $Z=600$ carboxyl surface groups, inside 
	suspension containing $300$ mM of 1:1 electrolyte  for  various pH values, 
	see Fig. \ref{fgd}. The density of sites on colloidal surface is the same as found in experimental systems~\cite{behrens2000charging}.
	We see a  good agreement between simulations and theory.  The deviations appear for small pH values, when most of  surface groups are protonated and small number of deprotonated groups can not be accurately described by the  continuum mean-field theory developed here.  Nevertheless, even under these extreme conditions, the agreement between theory and simulations is quite reasonable.

	\section{Transcendental Approximation}
	
	Solution of differential Eq.~\ref{mpb} with charge regulation boundary condition is numerically involved.  Often, one does not need the full ionic distribution, but only the effective colloidal charge at very low volume fractions.    In this case the calculation can be significantly simplified by letting the radius of WS become very large,  $R\rightarrow \infty$.  Under these conditions the mean electrostatic potential  at contact,   $\phi_c$, at distance $r_{ion}$ from colloidal surface  of a {\it uniformly} charged particle  can be related with the effective charge by~\cite{coldis}
	\begin{equation}\label{PB}
	Z_{eff}=-\frac{2 \kappa (a+r_{ion})^2}{ \lambda_B}\left[\sinh\left(\frac{\beta\phi_c}{2}\right)+\frac{2}{\kappa (a+r_{ion})}\tanh\left(\frac{\beta\phi_c}{4}\right)\right] \,.
	\end{equation}
	The first term on the right hand side is the usual relation between the surface charge density and surface potential for a planar PB equation,  while the second term is the leading curvature correction~\cite{coldis}.  Within the ion free layer, $a<r<a+r_{ion}$,
	the mean electrostatic potential is then
	\begin{equation}\label{free1}
	\beta \phi(r)=-Z_{eff} \lambda_B \left[\frac{1}{r}-\frac{1}{a+r_{ion}}\right]+\phi_c.
	\end{equation}
	The electrostatic potential on the surface of colloidal particles with a uniform surface charge  density $\sigma$ is then  
	\begin{equation}\label{free2}
	\beta \phi_0=\beta \phi(a)=- \frac{Z_{eff} \lambda_B r_{ion}}{a (a+r_{ion})}+\phi_c.
	\end{equation}
	Substituting equations (\ref{PB}), (\ref{free2}) and (\ref{EOCP2}) into the charge regulation equation,
	\begin{equation} \label{zeff2}
	Z_{eff}=\frac{Z}{1+ K c_a e^{-\beta(\phi_0+\phi_{disc}-\mu^{ex}-\mu_{sol})}},
	\end{equation} 
	we obtain a self-consistent equation for the contact potential $\phi_c$, from which the effective charge can be calculated directly using Eq. (\ref{PB}).  This procedure is much simpler than solving the spherical PB equation with charge regulation boundary condition.  {\color{black} On the other hand, it does not allow us to study the dependence of the effective charge on colloidal concentration or calculate the ionic density profiles.}
	For large WS cells, however, we see an excellent agreement between the colloidal charges calculated using the transcendental approximation and the full theory, see the colloidal charges calculated for a particle
	of radius 80~\AA\ with surface groups with intrinsic pKa$=5.4$ in a suspension containing  300 mM of 1:1 electrolyte at different pHs, presented in Table 1.
	
	\begin{table} 
		\centering
		\begin{tabular}{ccc}
			\hline
			pH	& Full Theory & Transcendental Approximation \\
			\hline
			4 & -6.84 & -6.80  \\
			\hline
			5 &  -33.151 & -32.592  \\
			\hline
			6 & -80.777 & -79.119 \\
			\hline
			7 &  -112.68 & -111.9 \\
			\hline
			8 &  -118.7 & -118.68  \\
			\hline
		\end{tabular}
		\caption{\label{ table }Comparison of the surface charge density in mC/m$^2$ obtained using numerical solution of the spherical non-linear PB equation with CR boundary condition inside a WS cell of $R=140$~\AA\  with the transcendental approximation method.} 
	\end{table}
	
\begin{figure}[H]
		\centering
		\includegraphics[width=0.7\linewidth]{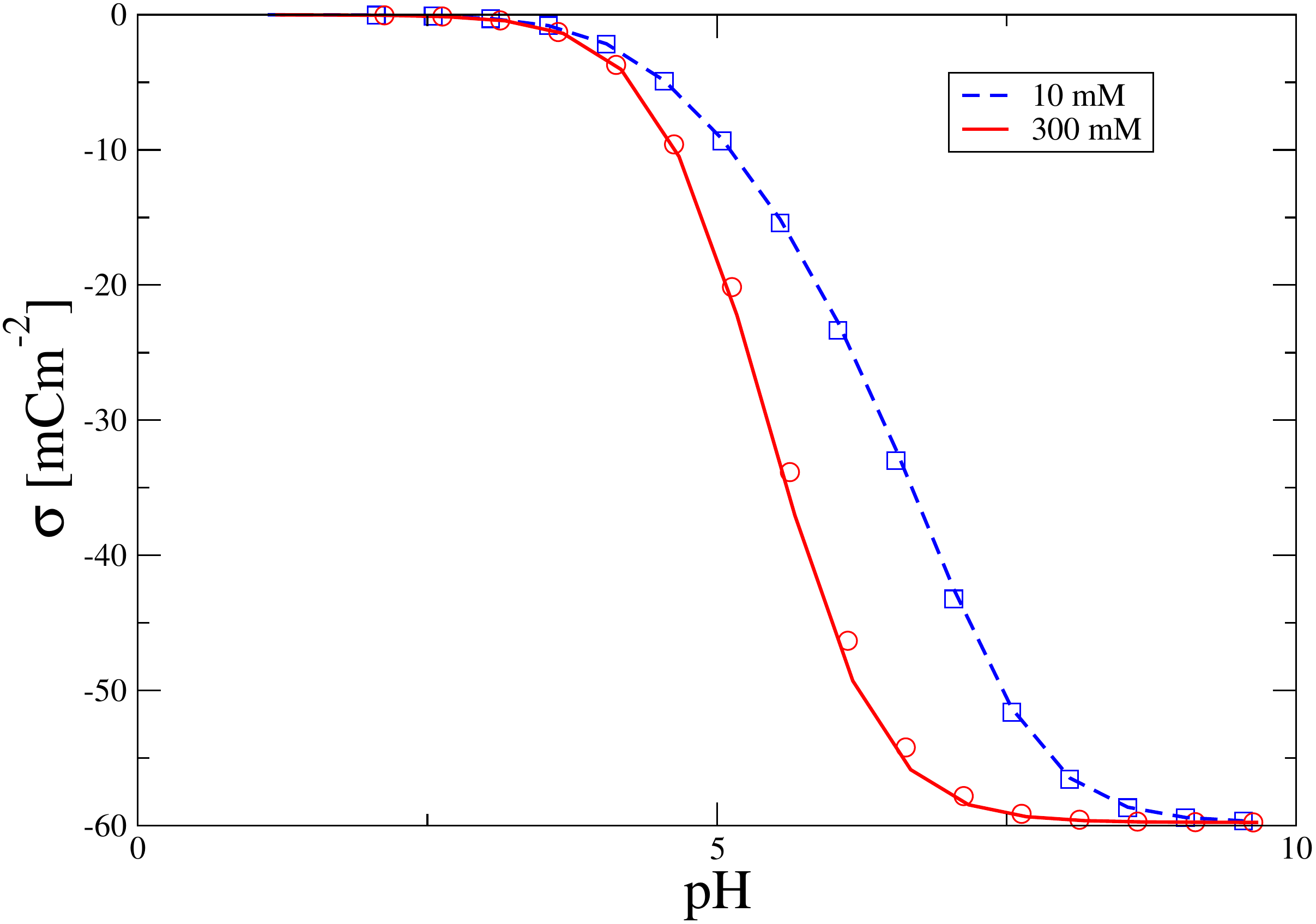}
		\caption{Titration curve for colloidal  particle with $-Zq=-59.6$~mC/m$^{2}$.   The theoretical curves are calculated using transcendental approximation equations.  Solid curve is for suspension containing 300 mM salt and the dashed curve is for 10 mM salt solution. {\color{black} The symbols are simulation results.}}
		\label{fig4}
	\end{figure}

\begin{figure}[H]
		\centering
		\includegraphics[width=0.7\linewidth]{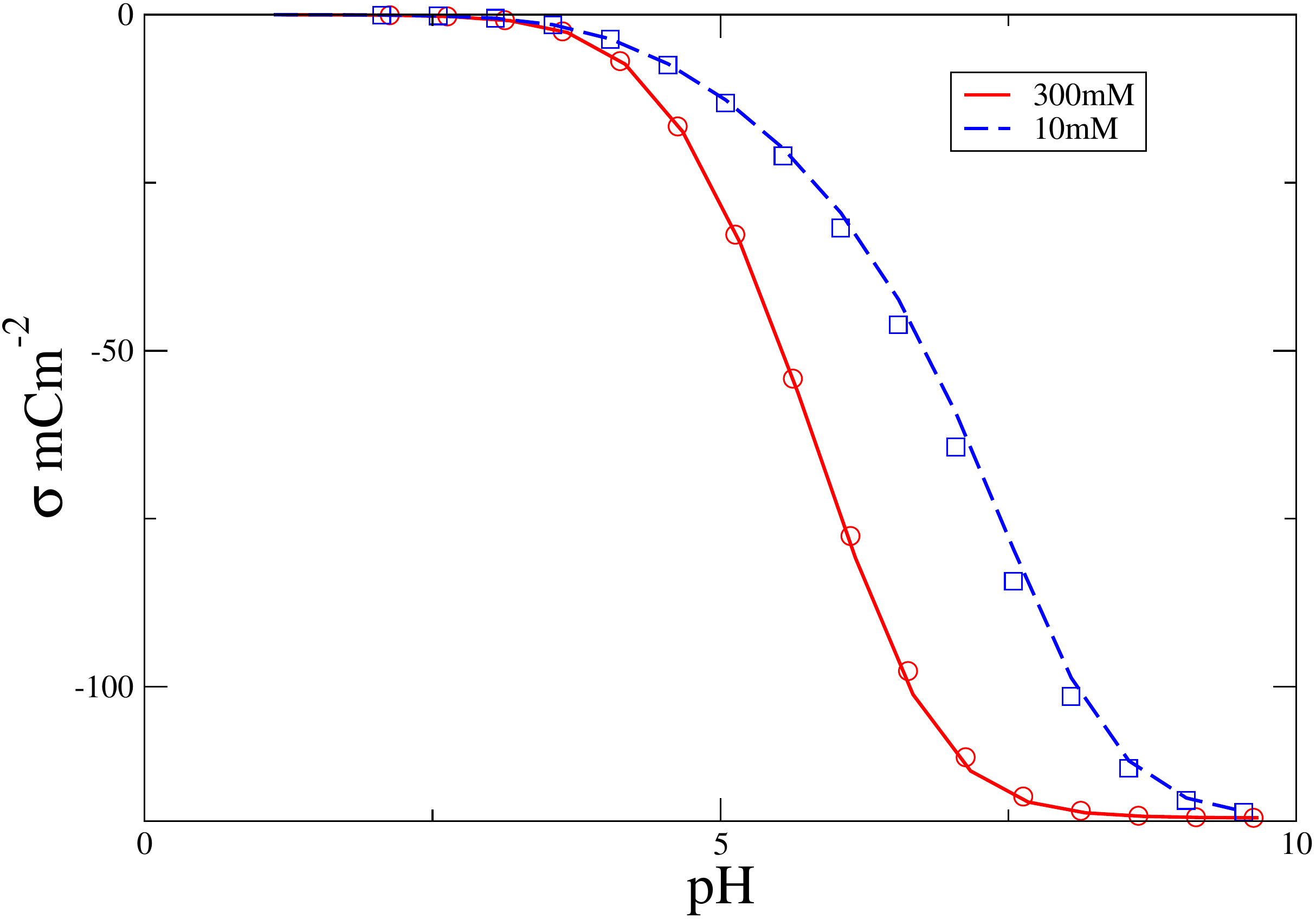}
		\caption{The comparison of titration curve obtained from simulation and theory. The number of acid groups is such that $-Zq=-119.4$~mC/m$^{2}$.   The solid 
			and dashed line are for the salt concentrations 300 mM and 10 mM, respectively.  {\color{black} The symbols represent simulation data points.}}
		\label{fig5}
	\end{figure}

\begin{figure}[H]
		\centering
		\includegraphics[width=0.7\linewidth]{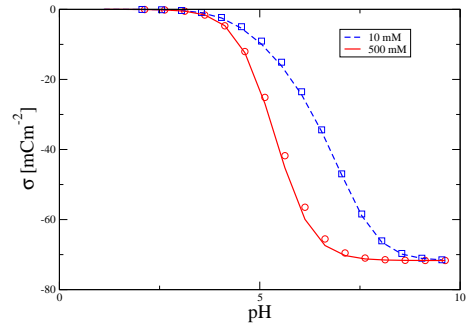}
		\caption{Titration of a colloidal particle with  $-Zq=-71.59$~mC/m$^{2}$. The solid and dashed line are for the salt concentrations 
			500 mM and 10 mM, respectively. {\color{black} The symbols represent simulation data and curves are the theory.}}
		\label{fig6}
	\end{figure}

\begin{figure}[H]
		\centering
		\includegraphics[width=0.7\linewidth]{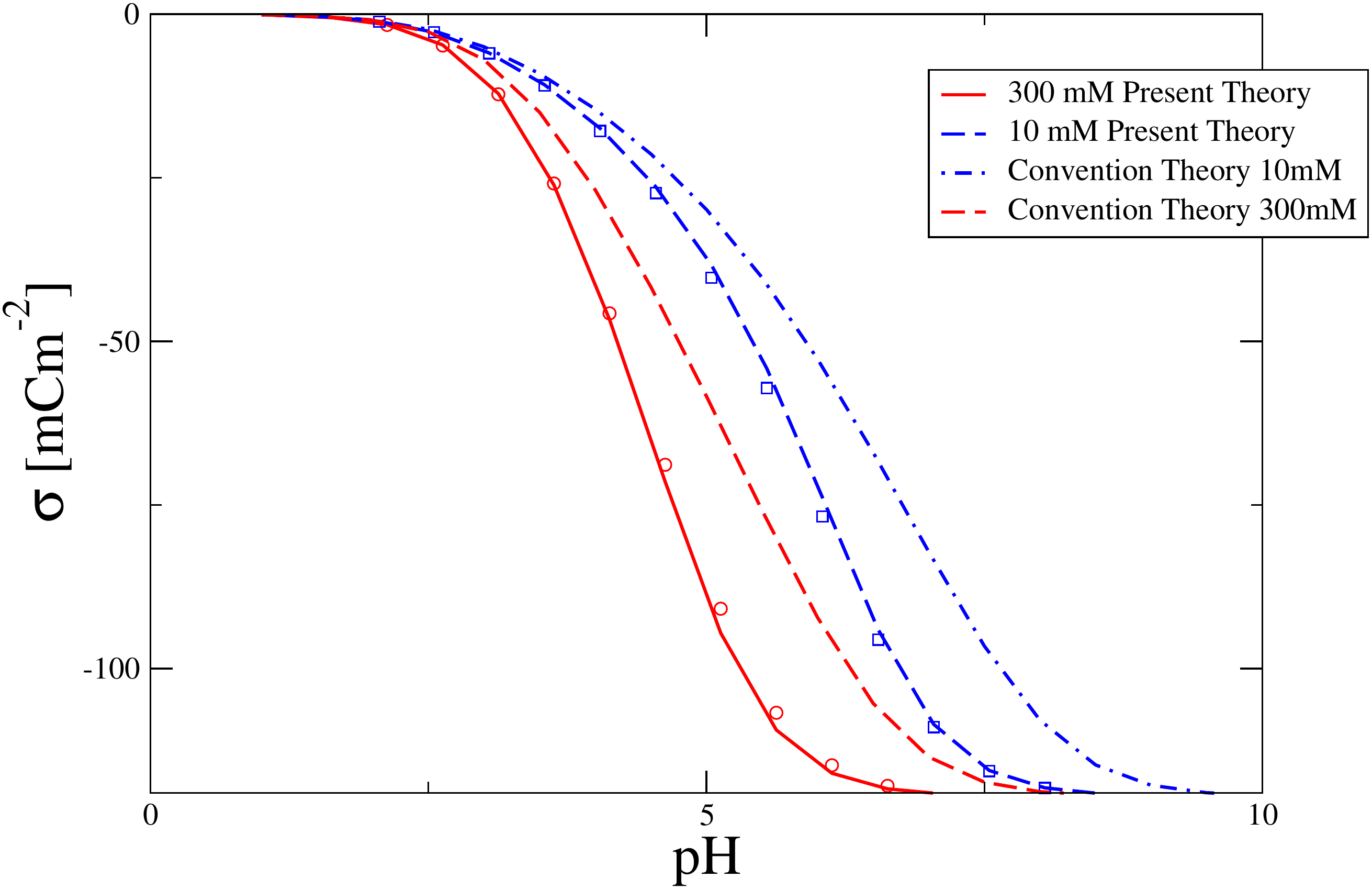}
		\caption{{\color{black} Titration of a colloidal particle of surface charge density $-71.6$~mC/m$^{2}$ with surface groups of intrinsic pK$_a =4.09$.  The symbols represent simulation data points. The curves correspond to the results of the present theory and to ``conventional'' titration, in which the discreteness and the correlational effects are neglected.}}
		\label{fig7}
	\end{figure}

\begin{figure}[H]
		\centering
		\includegraphics[width=0.7\linewidth]{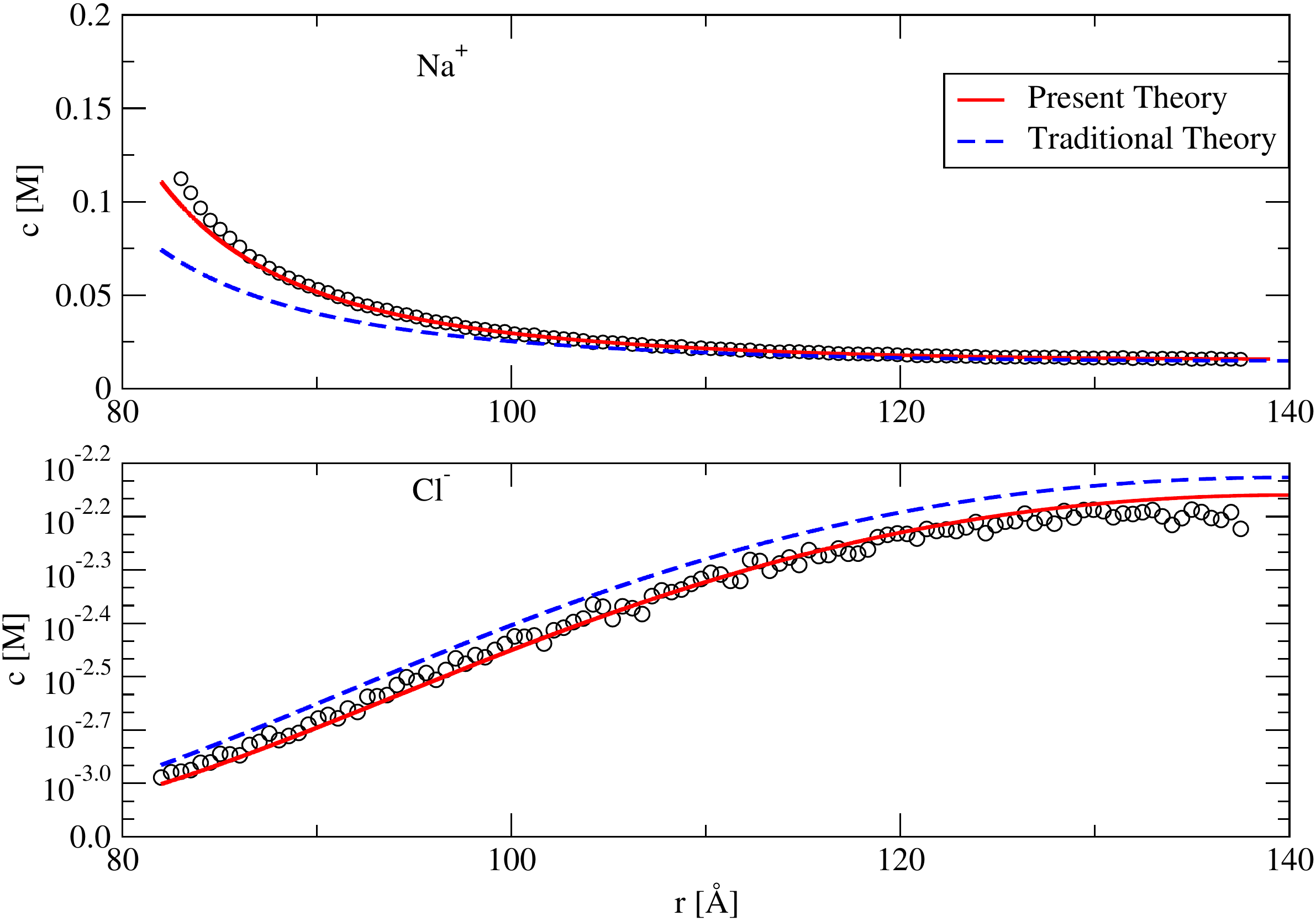}
		\caption{{\color{black}The ionic density profiles around a colloidal particle predicted by the present theory and by the conventional titration model in which discreteness and correlational effects are neglected.}}
		\label{last}
	\end{figure}

	As is demonstrated in Table 1,  there is a very good agreement between $\sigma$ obtained using the numerical solution of  spherical PB with CR  boundary condition and the one calculated using the approximate equation (\ref{PB}).   We can now use the  approximate equation (\ref{PB}) to efficiently calculate the titration curves of colloidal particles in suspensions of low volume fractions.  
	We start with a colloidal particle with $-Zq=-59.6$ mC/m$^{2}$ in a suspensions with either 10 mM or 300 mM of 1:1 electrolyte, see Fig.~\ref{fig4} 
	\vspace{1cm}
	We see an excellent agreement between theory and simulations, without any adjustable parameters.    We next study more highly charged particle with $-Zq=-119.4$~mC/m$^{2}$, see Fig.~\ref{fig5}
	Here once again, theory and simulations show good agreement.  The agreement persists even for higher salt concentrations, as can be seen 
	in Fig. \ref{fig6} showing the titration curves for salt concentration of 500 mM and 10 mM of particles with  $-Zq=-71.6$~mC/m$^{2}$. 
	{\color{black} To make sure that the good agreement observed between simulations and theory is not due to the specific value of $K_a$,  we next study colloidal particle with stronger acidic surface groups of $K_a=8.1 \times 10^{-5}$~M, pKa$=4.09$.  Furthermore,  to clearly see the effect of discreteness of surface charge groups and of electrostatic correlations on CR,  in addition to the present theory we also present the ``conventional"  titration curves in which these effects are neglected~\cite{nishio1994monte,borkideal}, i.e. $\mu^{ex} =\phi_{dis}=\mu_{sol}=0 $.   As can be seen from Fig.~\ref{fig7}, the present theory once again agrees very well with the simulations, while the conventional titration curves show strong deviations.}
	{\color{black} We next solve the full non-linear PB equation with our CR boundary conditions to calculate the ionic density profiles around this colloidal particles at pH=6 and 10 mM NaCl.  The theory once again shows a good agreement with simulations, 
		Fig.~\ref{last}.  On the other hand  when the discreteness and solvation energies are neglected, one sees strong deviations in ionic density profiles.}  
	
	
	{\color{black} In a recent paper~\cite{landsgesell2019simulations} authors argued that pH-pK$_a$ is a ``universal" parameter, namely that one will obtains identical number of protonated groups for systems with different pH and pK$_a$, as long as pH-pK$_a$$=$constant. Our theory shows that this is not the case, even in the absence of salt. To demonstrate this we study
		colloidal particles of radius 103.3~\AA~ with $Z=600$ surface groups, inside a WS cell of radius 140~\AA~.  The concentration of salt in the reservoir is set to zero.  First we fix the intrinsic pK$_a$ of surface groups to pK$_a=2.5$  and acidity to pH$=1$, so that pH$-$pK$_a$$=-1.5$. In this case our theory predicts that colloidal surface charge density will be  $-3.2$~mC/m$^2$. We then change the intrinsic pK$_a$ of surface groups to pK$_a=7.5$ and pH$=6$, so that again pH$-$pK$_a$$=-1.5$. For this system the theory predicts surface charge density of $-0.035$~mC/m$^2$. Clearly both are different, even though both systems have pH$-$pK$_a$$=-1.5$.  To confirm the predictions of the theory we ran rMC simulations.  The simulations yield surface charge densities of  $-2.9$~mC/m$^2$ and $-0.1$~mC/m$^2$, for the two cases respectively, in agreement with theory predictions.  Clearly pH-pK$_a$ is not a ``universal" parameter, contrary to the claims made in the reference~\cite{landsgesell2019simulations}.}
	
	{\color{black}{Finally, we note that within the present theory the classical Henderson-Hasselbalch (HH)
			equation  --- much used  in biochemistry and analytical chemistry to relate the value of pK$_a$ with the pH when half the surface groups are protonated, pH$_{1/2}$ ---  is modified to:
			pK$_a$ = pH$_{1/2}$ + $\beta q \varphi_{HH}   \log_{10}(\mathrm{e})$, 
			where $\mathrm{e}$ is the Euler number and 
			$\varphi_{HH}=\phi_{0} +\phi_{dis}-\mu_{sol}$ is the electrostatic potential at the center of an adsorption site minus the electrostatic solvation free energy of a deprotonated site. }}

	\section{Conclusion} 
	We have presented a theory which enables us to accurately calculate the surface charge of colloidal particles with uniformly distributed weak acid surface groups in solutions of various  pH and 1:1 electrolyte concentration. The theory accounts for the shift of solution pH due to the presence of electrolyte.  It also accounts  self-consistently for the electrostatic potential produced by the discrete  deprotonated surface groups.  To examine the accuracy of the  theory we  have performed extensive rMC simulations, which show excellent agreement between theory and simulations for all system parameters explored in the present paper. {\color{black} We have also used the theory developed in the present paper to demonstrate that contrary to recent suggestions~\cite{landsgesell2019simulations} pH$-$pK$_a$ is not a universal parameter. 
{\color{black} The theoretical approach to account for discreteness and solvation effects introduced in the present paper can also be included within the  MUltiSIte Complexation (MUSIC) model of Hiemstra {\it{et al.}}\cite{hiemstra1989multisite} used to study metal oxide surfaces, this  will be the topic of future work.}		
		
		Finally, it is well known that multivalent ions, such as \ch{Ca^{++}}, interact strongly with carboxylate.  In the future work we will attempt to extend the present theory to suspensions containing \ch{CaCl2} salt.  In that case, however,  presence of multivalent ions will lead to electrostatic correlations even in the bulk electrolyte.  This may require going beyond the PB equation and using classical density functional theory instead. 
	
	\section{Acknowledgments}
	
	This work was partially supported by the CNPq, the CAPES, and the National Institute of Science and Technology Complex Fluids INCT-FCx.  D.F. acknowledges financial support from the FONDECYT through Grant No. 1201192.

	\bibliography{ref}


\end{document}